\documentclass[12pt,letterpaper,preprint]{article}
\usepackage[top=1.4in,bottom=1.4in,left=0.80in,right=0.80in]{geometry}
\usepackage{graphicx}
\usepackage{pdfpages} % include pdfs
\usepackage{pgffor} % for loops

\usepackage{xcolor}
\usepackage{amsmath}
\usepackage{comment}
\usepackage{textcomp}
\usepackage[utf8]{inputenc}

\usepackage{authblk}
\usepackage[super,comma,compress]{natbib}

\usepackage{xcolor}
\DeclareUnicodeCharacter{2009}
\date{}
\begin{document}
\title{Specific yet transient bonds between anisotropic colloids}

\author{M Mayarani${^{1,2\,\ddag*}}$, Justine Laurent$^{3}$ Martin Lenz${^{1,2}}$, Olivia du Roure${^{1\,*}}$ and Julien Heuvingh${^{1}}$}
\affil{${^1}$~PMMH, CNRS, ESPCI Paris, PSL University, Sorbonne Université, Université
Paris-Cité, 75005, Paris, France \newline ${^2}$~Université Paris-Saclay, CNRS, LPTMS, 91405, Orsay, France.
\newline
${^3}$~Gulliver, CNRS, ESPCI Paris, PSL University, Sorbonne Université, 75005, Paris, France.
\newline ${^\ddag}$~Present address: Department of Physics, Indian Institute of Technology Palakkad, India, 678623.
Correspondence emails:\,mayarani@iitpkd.ac.in, olivia.duroure@espci.psl.eu}

\maketitle

\begin{abstract}
Self-assembly of colloidal particles is a promising avenue to control the shape and dynamics of larger aggregates. However, achieving the necessary fine control over the dynamics and specificity of the bonds between such particles remains a challenge. Here we demonstrate such control in bonds mediated by depletion interactions between anisotropic colloids that we 3D-print in the shape of half disks with sub-micron resolution. When brought together by diffusion, the particles interact in different configurations but the interaction through the flat faces is by far the longest-lasting. All bonds are flexible and transient, and we demonstrate control over their life time through the depletant concentration in  quantitative agreement with a simple physical model.
This basic design could be extended to manufacture particles with multiple binding sites to engineer directional assembly with multiple particles
\end{abstract}

\section{Introduction} 
 Colloidal self-assembly, leading to the formation of complex  hierarchical end products is key to
 understand fundamental processes such as glass transition\cite{weeks2000three,weeks2002subdiffusion}, crystallization\cite{wang2015crystallization,pusey1986phase,anderson2002insights}, polymerization\cite{mcmullen2018freely,luo2017polymerization} etc. It also possesses application prospects in various scenarios such as preparation of photonic crystals\cite{cai2021colloidal}, chemical sensing\cite{zhang20112}, biological applications\cite{kuo2013accelerated} and many more\cite{li2021colloidal,ge2011responsive}. At the heart of designing and organizing complex structures through self-assembly, lies the control on individual colloidal design and the mastery on their interactions. In various experimental attempts, researchers have demonstrated capability to tailor colloidal units that spontaneously self-assemble into intended structures based on chemical\cite{he2021colloidal}, geometrical\cite{sacanna_engineering_2013,sacanna2010lock,tigges2016hierarchical}, or physical cues\cite{bharti2016multidirectional,van2004colloids,dobnikar2013emergent}. Highly anisotropic colloidal interactions leading to the formation of directional bonds are achieved through altering the geometry, roughness and/or surface properties of colloids\cite{li2020colloidal}. However, good control over the formation, transient nature, life-time and dissociation of such bonds and their quantitative correspondence to theoretical predictions are still elusive. 

To build predictable aggregates from scratch, utilizing the principles of self-assembly, the building blocks and interactions should be carefully crafted to render the bonds (i) selective; that favor one type of assembly over another, (ii)~reversible; that allow reorganizations of the structure, which helps to avoid  energetically unfavorable energetic traps and (iii) flexible; to allow for compensation of any defect in the manufacture of the individual particles. Our goal in this work is to achieve a   good control over these aspects of colloidal self-assembly by  micro-printing simple half-disk like particle  and inducing short-ranged attractive interaction through depletion of polymers.

Depletion interactions are ideally suited to induce colloidal self-assembly. They can indeed be controlled independently of the colloid fabrication process and have been used to assemble lock-and-key particles through shape complementarity~\cite{asakura1954interaction,sacanna2010lock}. Depletion interactions take place between colloids in presence of non-adsorbing polymer:
As colloids come into contact, the volume between them becomes inaccessible to the polymer and it is favorable in terms of entropy to push the colloids into close contact.~\cite{colon2015binding}. For two colloids in  contact over an area $A$, the binding free energy is proportional to $c A \delta$, with  $c$ being the depletant concentration and   $\delta$ the range of interaction, which is proportional to the radius of gyration of the polymer.
Better fitting colloids with a larger area $A$ benefit from stronger interaction, which accounts for the efficient binding of lock-and-key colloidal designs. The magnitude of depletion interactions has been experimentally verified for small colloids and non-ionic polymers through surface force apparatus measurements~\cite{ruths1996depletion,kuhl1998part}, optical trapping~\cite{crocker1999entropic} and total internal reflection microscopy~\cite{rudhardt1998direct,edwards2012depletion}.
 One promising avenue to manufacture colloids with designed shapes, in addition to chemical synthesis and DNA origami\cite{tigges20163d}, consists of 3D printing them on a substrate\cite{brown2000fabricating,merkel2010scalable,hernandez2007colloidal,moon2007high,jang2007shape}.
 Two photon laser printing has recently been used to fabricate self-assembling colloids \cite{tigges2016hierarchical,doherty2020catalytically}, offering a large design flexibility at the price of a lower throughput as compared to chemical routes. While the self-assembly of such laser-printed colloids through depletion interactions has previously been demonstrated~\cite{tigges_hierarchical_2016}, the transient, re-configurable bonds required for faithful large-scale complex self-assembly have not yet been achieved in this setting. 
 
  Here we demonstrate such properties in Brownian colloids produced through direct laser writing based on two-photon polymerization. As described in Sec.~\ref{sec:methods}, we print colloids on a sacrificial polymer layer and observe the interactions between the particles locally~\cite{HernandezAlphabet, Tavacoli2013, Saraswat2020}. Their shape is semi-circular, which implies a stronger interaction through their flat faces than their curved faces. Depletion interactions are induced through the addition of polyethylene glycol chains. In Sec.~\ref{sec:results}, we characterize the formation of these bonds, their selectivity and fluctuation dynamics as well as their eventual rupture under the influence of thermal fluctuations as a function of the depletant concentration. We anticipate that our approach will constitute a versatile platform to further engineer complex self-assembling systems.

\section{\label{sec:methods}Methods}
To obtain Brownian colloids with controlled shapes and number density, we 3D print the colloidal particles on a sacrificial layer of poly acrylic acid(PAA) (Sec.\ref{3D-printing on paa}). Once printed, we liberate the colloids by dissolving the layer, which allows them to diffuse and interact \emph{in situ}, and control their interactions with depletion forces (Sec.\ref{particle detatchment} ).

     \begin{figure}[t]
\centering
  \includegraphics[height=5cm]{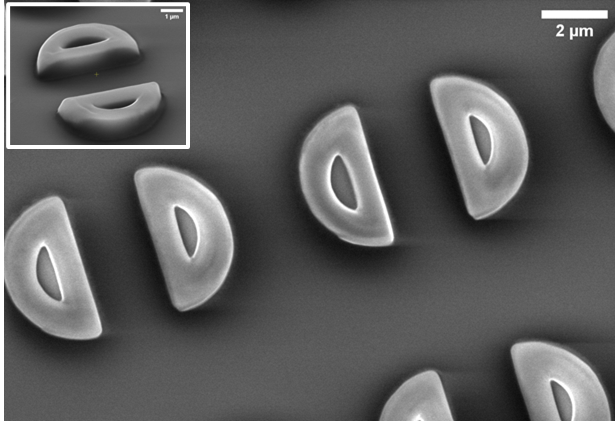}
  \caption{Electron microscopy image showing the printed colloids, captured at an accelerating voltage of 15\,kV. Inset: tilted image.}
  \label{fgr:SEM_of_sample}
  
\end{figure}

\subsection{\label{3D-printing on paa}3D-printing  of colloids}
Colloidal particles of half-disk shape are fabricated using 3D printing technique 
based on two-photon polymerization. 
To fabricate colloidal particles
a sketch of the desired particle geometry is first made using a 3D designing software, Autodesk Inventor professional. Our colloids are designed to be semi-circular in shape with a diameter of $5\mu$m and a height of $1\mu$m. The particles are designed to be flat, in order to keep them parallel to the substrate during self-assembly, and to minimize particle flipping. The design is then loaded into the nanowrite software linked to the 3D printer, Photonic professional GT from Nanoscribe, Germany. The 3D design is vertically sliced into parallel planes at fixed distances using the nanowrite software. Since the approximate height and diameter of a single voxel resulting from the tight focusing of laser onto the resist is 0.8 $\mu m$ and 0.3 $\mu m$ respectively, we maintain the  vertical slicing and horizontal hatching distances at 0.2 $\mu m$ during particle printing. This ensures optimal overlap between the neighboring voxels 

Colloids are printed in the conventional mode of direct-laser writing where the laser is focused through a thin glass substrate onto a photosensitive material\,(photo-resist). A circular cover glass of 30 mm diameter and 1.5 mm thickness is used as the substrate. After thorough oxygen plasma cleaning, we first coat the  glass substrate  with a thin layer of PAA using a 20mg/ml solution at 3000rpm for 30 seconds. A drop of IP-L photo-resist (from Nanoscribe GmbH, Germany) is placed on top of the PAA layer. The substrate is then loaded onto the 3D printer and the laser is focused onto the resist from underneath the glass substrate using a 63x oil-immersion objective (see Supporting Information Fig.~S1 for a schematic). The laser then writes the 3D structures onto the photoresist which reticulates and solidifies. The unreacted photoresist is washed off using propylene glycol methyl ether acetate (Sigma Aldrich) leaving the printed colloidal particles on the substrate as the sacrificial PAA layer is insoluble in the developer solvent. A scanning electron microscopy\,(SEM) image of the printed particles is shown in Fig.~\ref{fgr:SEM_of_sample}. The printed semi-circular particles have a diameter of 4.60$\pm $0.05$\,\mu m$ as measured on SEM images and a height of 0.82 $\pm$0.06$\,\mu m$ as inferred from optical microscopy measurements (detailed in the supporting information (see Supporting Information Fig.~S2)). The hole in the middle of the colloids enables their easy detection and analysis in particular to measure the centroid and orientation (see Supporting Information Fig.~S3). 
 \subsection{\label{particle detatchment}Detachment of the particles from the printing substrate}
 \paragraph*{} The printed particles are liberated from the substrate by dissolving the PAA sacrificial layer in a depletant solution that contains different chemicals dissolved in water: To induce depletion interactions and shape-selective binding between the printed colloids, we introduce polyethylene glycol (PEG) (MW 600\,kDa, from Sigma Aldrich) as the depletant. To prevent unfavorable and irreversible binding between colloids we use a non-ionic surfactant, tergitol at 20 ppm (from Sigma Aldrich), and salt at 50 mM (NaCl) to screen electrostatic repulsion between the printed colloids by decreasing the Debye screening length of the system. 
   
Before dissolving the PAA layer to release the particles, we make a small chamber around the printed colloids to contain the solution and prevent fluid flow. A rubber `O-ring' is first fixed to the glass substrate around the printed colloids by using a thin layer of silicon oil. About 20$\mu l$ of depletant solution is carefully placed inside the chamber, which is then sealed with a glass cover slip to avoid evaporation (see Supporting Information Fig.~S4 for a schematic).  The colloids are observed using an optical microscope from Carl Zeiss in bright field mode under 100x magnification using an oil-immersion objective. Time-lapse movies are acquired through a Michrome 6 CMOS camera (Keyence, France).
    
    \section{\label{sec:results}Results}
    After dissolution of the sacrificial layer, the colloids start to diffuse (Sec.~\ref{individual particle diffusion}) and upon contacting one another form bound pairs due to depletion interaction whose lifetimes (Sec.~\ref{observe bond formation}), fluctuations (Sec.~\ref{sec:bond_fluctuations}) and breaking behaviour (Sec.~\ref{sec:bond breaking}) we characterize below alongside their dependence on the depletant concentration (Sec.~\ref{sec:concentration dependance of bonds}).
    
    \subsection{\label{individual particle diffusion}Diffusion of individual particles}
    As the depletant solution is administered into the chamber, the PAA sacrificial layer dissolves, setting the printed particles free to diffuse along the glass surface. Gently placing the depletant solution liberates the colloids without flipping them. The liberated particles remain close to the bottom of the chamber due to their higher density compared to water. As we show below, bonds form between the colloids and detach. In our movies that are acquired at least ten minutes after dissolution of the PAA layer, we can safely consider that the initial orientational order is completely randomized by thermal fluctuations (see Supporting Information figure S5).
    The geometric center of diffusing particles is tracked over time, to monitor their trajectories. By calculating the mean square displacement of the diffusing particles, we calculate the diffusion coefficient of the individual colloidal particles to be equal to $0.044\pm0.002\,\mu \text{m}^2/\text{s}$ (see Supporting Information Fig.~S6). This diffusion is sufficient to cause several contacting events between neighboring colloids in the typical course of our experiments.

    \subsection{{\label{observe bond formation}Direct observation of transient bond formation and breakage}}
\begin{figure*}[t]
\centering
    \includegraphics[width=0.9\textwidth]{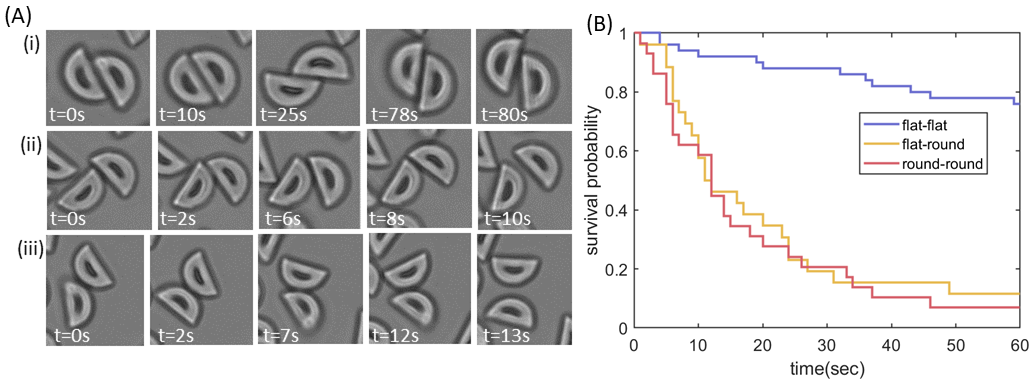}
  \caption{\textbf{The life time of a bond depends on its geometry.} (A)~Time lapse of bond formation, evolution and breaking between a pair of colloids in (i) flat-flat (ii) flat-round and (iii) round-round configuration. Note that the timescales (time is indicated on every image) vary a lot between the first row and the two others. (B)~Survival probability of the three different types of bonds formed in the system, at a depletant concentration of 0.01 mg/ml PEG. The measurement is carried-out for a minimum of 30 pairs in each configuration. }
  \label{fgr:microscopy_panels_life_time}
\end{figure*}

Two colloids coming into contact can do so in three different configurations: their respective flat faces may come together, or their round ones, or they may incur a mixed flat-round contact. Fig.~\ref{fgr:microscopy_panels_life_time}(A) shows the formation, temporal evolution, and breaking of the three types of bonds observed in our system at a depletant concentration of 0.01\,mg/ml. The time at which bond formation takes place is denoted as t=0. In all three cases, individual colloids     fluctuate with respect to each other. The flat-flat bonds are longest-lived, indicating that the bonds between colloids are shape-selective. 

To further establish the shape specificity of depletion interaction in our system, we observe several bound pairs of each category over 60 seconds and identify the time of bond breakage. Fig.~\ref{fgr:microscopy_panels_life_time} (B) shows the survival probability of each of the three types of bonds at 0.01\,mg/ml PEG. Flat-flat bonds have a considerably higher survival probability compared to flat-round or round-round configurations, confirming that our design allows bond specificity through colloid shape to be programmed.

\subsection{\label{sec:bond_fluctuations}{Bond fluctuations}}
  
\begin{figure*}[t]
\centering
    \includegraphics[width=0.9\textwidth]{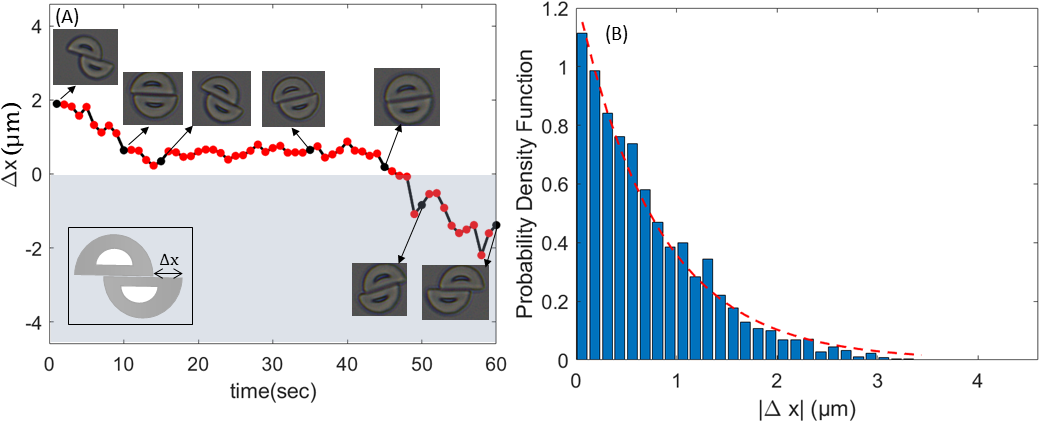}
  \caption{\textbf{Bond fluctuations are consistent with those of an equilibrium steady state.}(A)~Temporal evolution of the offset $\Delta x$ schematized in the inset between the adjacent edges of  colloids showing the fluctuation of a bond in flat-flat configuration. (B)~Probability density function corresponding to the occurrence of $|\Delta x|$ values during fluctuation of bonds in flat-flat configuration formed by the depletion-assisted association of semi-circular colloids at a depletant concentration of 0.01mg/ml, with exponential fit (red dotted line) superimposed.}
  \label{fgr:positive_negative}
\end{figure*}

To investigate the fluctuations of a single flat-flat bond, we plot the offset $\Delta x$ between the adjacent edges of the two colloids  as a function of time in Fig.~\ref{fgr:positive_negative}(A). For each micrograph in our time series, we obtain the value of this offset by
 projecting the centroid of one of the ellipses to the major axis of the second ellipse, and then calculating the distance between the point of projection and the centroid of the second ellipse. When the flat faces of the two colloids are perfectly aligned with each other, the offset value is zero. The offset takes positive or negative values depending on the relative positions of two colloids while they fluctuate. The pair of colloids represented in Fig.~\ref{fgr:positive_negative}(A) explores a range of $\Delta x$ values ranging from $-2\mu$m to $+2\mu$m  within the observation time of 60 seconds, and we show snapshots of the state of the bond at the times indicated by black dots. Despite this highly flexible nature, the bound pair spends most of its time in configurations close to $\Delta x=0$, which have the lowest depletion free energy

To determine whether the fluctuations of the bond are consistent with our simplified picture of two perfectly flat faces constrained solely by depletion interactions, we follow $42$ flat-flat bonds over 1 minute with a frame rate of 1 per second and plot the probability density function of $|\Delta x|$. Equilibrium thermodynamics predicts that this probability distribution should take the form of a Boltzmann distribution $e^{- \Delta F/k_B T}$, with $\Delta F\propto |\Delta x|$ the loss of depletion free energy. The data shown in Fig.~\ref{fgr:positive_negative}(B) is consistent with this prediction. We fit the data  with an exponential of the form $p(\Delta x)=\lambda^{-1} e^{- |\Delta x|/\lambda}$, and obtain an estimate of $\lambda=0.80 \pm 0.04 \mu m$. Estimating the change in excluded volume as $\Delta V =  2 h \delta |\Delta x|$, where $\delta$ is the thickness of the depletion layer, we obtain a theoretical prediction of $\lambda_{th} = 2 h \delta c  =0.96 \pm 0.07 \mu m$, close to the measured value. The difference between prediction and measurement may be due to the friction between the two surfaces, which is unaccounted for here. This model thus suggests that bonds are mostly observed in configurations whose depletion free energy is within $k_BT$ of its minimum value. Consistent with this expectation, our bonds rarely display very large values for $|\Delta x|$ despite remaining dynamic.

 \subsection{\label{sec:bond breaking}{Bond breaking}} 
 While bond configurations with large values of $|\Delta x|$, are rare, we expect that they would be the most likely to break apart due to the small overlap of the half-disks. To assess the kinetic pathway leading to bond breakage, we measure the angle $\theta$ between the major axes of the elliptical fits of the two particles. When the particles are bound, the two flat faces stay parallel to each other, making an angle between the two ellipses close to $0^\circ$. However, sometimes the angle between the major axes abruptly increases (see Supporting Information Fig.~S7) and the particles separate. In this case, we record the corresponding bond breaking time, and  $|\Delta x|$ value. 
  
  \begin{figure*}[t]
\centering
  \includegraphics[height=6cm]{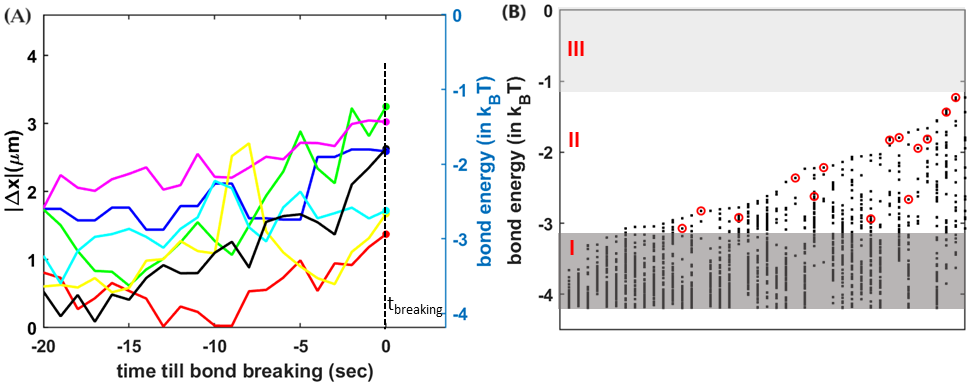}
  \caption{Bond breaking dynamics of flat-flat pairs: (A) Temporal evolution of $|\Delta x|$ recorded from 20 seconds prior to breakage of the bond till the breaking time, $t_{breaking}$ corresponding to 7 different flat-flat pairs formed at a depletant concentration of 0.01 mg/ml (left Y-axis) and the corresponding bond energies in units of $k_BT$ (right Y-axis). The $|\Delta x|$  at $t_{breaking}$ is denoted by filled circles for each pair. (B) Bond energy values explored by various pairs of colloids in flat-flat configuration at 0.01 mg/ml depletant concentration. The bond energies at breaking is indicated by red open circles. Out of the 42 bonds represented here, 14 bond breaking events are observed.}

  \label{fgr:breaking_bond_20sec}
\end{figure*}

We show the time evolution of the offset $|\Delta x|$ for 7 flat-flat bonds in Fig.~\ref{fgr:breaking_bond_20sec}(A), with time $0$ indicating their breakage. We observe that breakage tends to occurs for relatively high offsets. To confirm this observation, we summarize the behavior of 42 flat-flat pairs in Fig.~\ref{fgr:breaking_bond_20sec}(B). In this figure, each column shows the different values of $|\Delta x|$ (black dots) explored by a pair during the course of an experiment (one minute). The value at which the pair breaks, if it exists, is represented by a red circle. The pairs are ranked horizontally based on their highest $|\Delta x|$ value. 
The first observation of this figure reveals that a pair can explore offset values greater than the one at which it - or the others - breaks, reminding us that the phenomenon we are studying is stochastic because it is induced by thermal fluctuations. 
Second, three different regions corresponding to different offset ranges are visible. Although all pairs visit the low offset region\,(I), no bond breakages are observed there, which is consistent with an associated binding free energy that is large relative to the thermal energy (in excess of $3k_BT$) leading to a very favorable binding. All observed breakage events occur in region II, which corresponds to a binding free energy comprised between $1.2$ and $3k_BT$, low enough to be overcome by thermal fluctuations over the time scale of our observations. Finally none of our bonds explores the  high-offset region III, which would correspond to very weak bonds. In the next section, we investigate how these behaviours are affected as we alter the bond energy by changing the concentration of depletant.  

 \begin{figure*}[t]
\centering
    \includegraphics[width=1\textwidth]{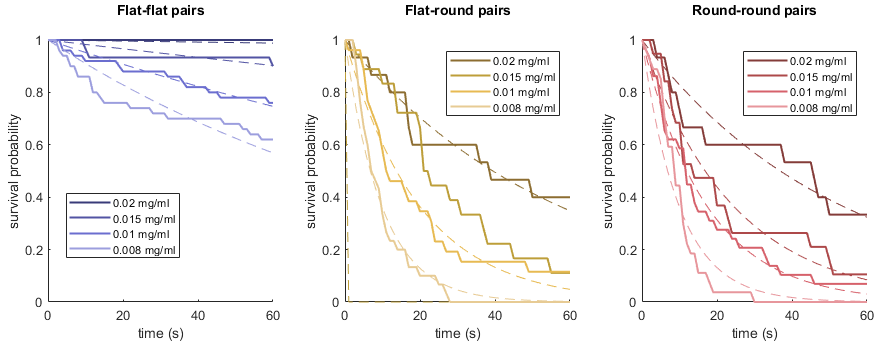}
  \caption{Dependence of survival probability of the three types of bonds in (i) flat-flat (ii)flat-round and (iii) round-round configurations for four different concentration of PEG depletant. Dashed lines: fit to obtain the survival times.}
  \label{fgr:life_time_with_concentration}
\end{figure*}

\subsection{\label{sec:concentration dependance of bonds}{Influence of the depletant concentration}}

We investigate the effect of depletant concentration on the decay and breaking behaviour of various bonds formed in our system. Firstly, we assess the survival probability of colloidal pairs in the three different bond configurations viz flat-flat, flat-round, round-round formed at various depletant concentrations (0.02 mg/ml, 0.015 mg/ml, 0.01 mg/ml and 0.008 mg/ml PEG) (see Fig.~\ref{fgr:life_time_with_concentration}). To obtain a quantitative estimation of the bond lifetime $\tau$, we fit the survival probablities with a decreasing exponential as a function of time  ($p=e^{-t/\tau}$). The survival probability decreases for each type of bonds as the strength of depletion interaction is lowered systematically. Consequently, with decrease in depletant concentration of depletant, the decay rate of bonds increases, indicating declining bond stability with decrease in the depletant concentration. This is true for all the three types of bonds observed in our system. Fig.~\ref{fgr:life_time_with_concentration} also re-emphasises the specificity of colloidal bindings, a feature already evidenced from Fig.~\ref{fgr:microscopy_panels_life_time} (B). At each depletant concentration, flat-round and round-round bonds exhibited relatively shorter life times compared to the corresponding flat-flat bonds, which are long lived and stable.

In a first approach, we assimilate bond breakage to the escape from a single potential well of depth $\Delta F$, where $\Delta F$ denotes the depletion free energy associated with a bond configuration. According to Kramers theory, the survival probability of the bond should then follow the type of exponential decay described above with a mean detachment time given by the Arrhenius law  $\tau = \tau_0  e^{\Delta F /k_BT}$, where $\tau_0$ is a constant typical diffusion time scale for the problem. $\Delta F$ is given by the product of the osmotic pressure and the change in excluded volume $\Delta V$ yielding $\Delta F = k_BT c \Delta V$. We thus obtain a simple prediction for the survival probability $\ln(\tau) = \ln(\tau_0) + c \Delta V $. We then fit the three curves of $\tau$ as a function of concentration with four parameters, $\Delta V$ for each configuration (flat-flat, flat-round and round-round) and the same pre-exponential constant $\tau_0$. These values are very close to a geometric estimate of the excluded volume change for each of the bond configurations (see Table \ref{table:1} and supplementary data for calculation).  
 \begin{figure}
\centering
    \includegraphics[width=0.45\textwidth]{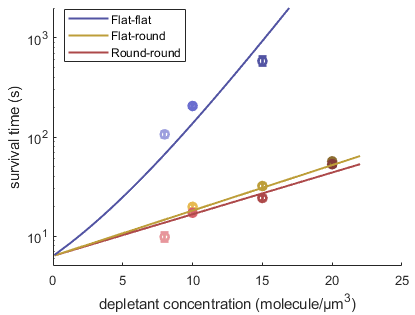}
  \caption{Survival time as a function of depletant concentration for the flat-flat, flat-round and round-round configurations, with associated fits of Arrhenius law, modified in the flat-flat case for reduced survival associated with sliding.}
  \label{fgr:Arhenius}
\end{figure}

Despite this good agreement, we speculate that the deviations of our fitted values from our geometrical estimates could stem from the flexibility of the flat-flat bonds. Indeed, as the flat surfaces of two bound colloids randomly slide off of a perfect alignment due to thermal fluctuations, they reduce the overlap of their excluded volume. This increases the rate of their detachment, and potentially offers a fast, ``slide-then-break'' kinetic pathway towards bond breakage. This hypothesis is supported by the observation of the flat-flat detachment scenario showing sliding at the 4 different depletant concentrations studied (SI Fig.~S9). Although detachment occurs at all offsets at the lower depletant concentration and is not observed at the highest concentration, at intermediate concentrations, detachment occurs only for the highest offsets. Plotting the detachment scenario with the calculated bond energy instead of the offset, a qualitative agreement is obtained with (SI Fig.~S9), where all flat-flat pair separations occurs between an energy of $-4 kT$ to $-1 kT$.
\begin{table*}[t]
\begin{center}
\begin{tabular}{|c || c c c|} 
 \hline
  & flat-flat & flat-round & round-round \\ [0.5ex] 
 \hline\hline
$\Delta V$ from geometry ($\mu \text{m}^3$) & $0.426 \pm 0.032$  & $0.089 \pm 0.007 $ & $0.063 \pm 0.005 $ \\ 
 \hline
$\Delta V$ from Arhenius ($\mu \text{m}^3$) & $0.338$ & $0.118$ & $0.109$ \\ 
 \hline
$\Delta V$ from Arhenius with sliding ($\mu \text{m}^3$) & $0.462$ & $0.106 $ & $0.097$ \\ [1ex] 
 \hline
 \end{tabular}
 \caption{Excluded volume change when two particles contact, calculated from the geometry of the particle and compared with estimates derived from fits of the survival time as a function of concentration.}
\label{table:1}
 \end{center}
\end{table*}

Beyond this qualitative observation, we aim to validate our slide-then-break hypothesis by modeling it quantitatively. For a shift $\Delta x$, the binding free energy between the colloids is reduced by an amount $2k_BTc h\delta|\Delta x|$. By assuming a simple Arrhenius kinetics for their detachment, we thus predict a detachment rate $k(\Delta x)=\tau_0^{-1}\exp[cV(\Delta x)]$, where the overlap between the colloids' depletion volumes is given by $\Delta V=2 h\delta (R-|\Delta x|)$. Using the probability $p(\Delta x)$ derived in Sec.~\ref{sec:bond_fluctuations}, we compute the mean colloid detachment rate $K=\int p(\Delta x)k(\Delta x)\,\text{d}\Delta x$. Defining the effective overlap $\Delta V_\text{eff}$ between the two colloids through  $K=\tau_0^{-1}\exp(c \Delta V_\text{eff})$, our calculation yields 
 \begin{equation}
     c \Delta V_\text{eff} = c V_M + \ln\left(\frac{1-e^{-c \Delta V_M}}{c \Delta V_M}\right),
 \end{equation}
where the maximum overlap is given by $\Delta V_M=\Delta V(\Delta x=0)$. We then fit again (see Fig.~\ref{fgr:Arhenius})
using this modified model the ensemble of the three variations of tau as a function of concentration, to obtain an estimation of $\Delta V_M$ for flat-flat, and a new estimate of $\Delta V$ for flat-round and round-round configuration.The results are shown in Table\ref{table:1} and show a much better agreement with the geometric estimates. This better agreement supports our slide-then-break kinetic model, and thus demonstrates that depletion forces can adequately explain the dynamics of aggregation of these micro-fabricated colloids and that other attractive forces, such as van der Waals interaction, only plays a minor role.

\vspace{-2mm}
\section{Discussion}
\paragraph*{}
Successfully self-assembling particles into a predetermined structure involves two challenges. On the one hand,  the target structure should be more stable than its competitors. On the other, it must be kinetically accessible. 
While 3D-printed microparticles offer a remarkable flexibility in achieving complex stable structures, kinetic accessibility is potentially problematic at their relatively large scale, where diffusion is much slower than in, \emph{e.g.}, DNA origami. This difficulty can however be offset by a fine control over the interactions between the particles, \emph{e.g.}, allowing off-target bonds to quickly detach, while even favorable ones are allowed to occasionally come off to allow the particles to optimize their large-scale arrangement. 

In this study, we have demonstrated an experimental strategy to achieve such control. We  implement reversible bonding with a life time directly controlled by the particles' shapes, demonstrating that 3D printing can be used to not only control an aggregate's morphology, but also its dynamics. Since the strength of our bonds is controlled both through the depletant concentration and the contact area between the  colloids, our design can straightfowardly be generalized to generate a system where different bonds have different lifetimes. This potentially opens the possibility to use 3D printing to design structures based on hierarchical self-assembly, which take advantage of the existence of several scales of bond life time and strength to reliably assemble complex structures.
 We also demonstrate that the spatial control afforded by 3D printing can be harnessed to generate flexible aggregates whose rigidity is controlled by easily controlled depletion interactions. While such sliding of flat colloid surfaces in the presence of depletion attraction have previously been observed with silica cubes\cite{rossi_cubic_2011}, ours is to our knowledge the first implementation of this effect in a 3D-printed self-assembled system. On larger scales, this sliding can in principle be controlled through the size of the depletant, an effect that has previously been used to select between different lattice organisations\cite{sacanna_engineering_2013}.

In addition to allow for the implementation of non-rigid self-assembled structures, we demonstrate that the flexibility of a bond has a direct influence over its lifetime through a slide-then-break mechanism, allowing one more lever to control the dynamics of colloidal self-assembly by taking advantage of the spatial control afforded by 3D-printing. While our study concentrates on the dynamics of a single inter-colloid bond, we anticipate that our design can easily be scaled up to generate particles with multiple binding sites. As this strategy opens the way to much more complex designs, flexibility could prove an asset in yet another way. Specifically, in such a context bond flexibility could compensate for imperfections in the colloids' shapes, by allowing, \emph{e.g.}, a ring of particles each carrying two bonds at an angle to one another to close even in cases where these angles are not exactly adjusted. We thus anticipate that the tool box developed here could dramatically open the range of possible designs for the self-assembly of 3D-printed objects, both in the quasi-two-dimensional setting considered here and in future 3-dimensional situations.
%\vspace{-5mm}
\section{Acknowledgments}
This work was supported by the \lq\lq
   Défi Auto-Organisation\rq\rq  of CNRS' mission of transverse and interdisciplinary initiatives and ANR grant\,(ANR-22-CE30-0024). ML was supported by ERC Starting grant 677532, ANR's Tremplin ERC grant ANR-21-CE11-0004-02 the Impulscience\textsuperscript{\tiny{\textregistered}} program of Fondation Bettencourt Schueller.

\bibliographystyle{unsrtnat}
\bibliography{ref}

\begin{thebibliography}{39}
\providecommand{\natexlab}[1]{#1}
\providecommand{\url}[1]{\texttt{#1}}
\expandafter\ifx\csname urlstyle\endcsname\relax
  \providecommand{\doi}[1]{doi: #1}\else
  \providecommand{\doi}{doi: \begingroup \urlstyle{rm}\Url}\fi

\bibitem[Weeks et~al.(2000)Weeks, Crocker, Levitt, Schofield, and Weitz]{weeks2000three}
Eric~R Weeks, John~C Crocker, Andrew~C Levitt, Andrew Schofield, and David~A Weitz.
\newblock Three-dimensional direct imaging of structural relaxation near the colloidal glass transition.
\newblock \emph{Science}, 287\penalty0 (5453):\penalty0 627--631, 2000.

\bibitem[Weeks and Weitz(2002)]{weeks2002subdiffusion}
Eric~R Weeks and David~A Weitz.
\newblock Subdiffusion and the cage effect studied near the colloidal glass transition.
\newblock \emph{Chemical physics}, 284\penalty0 (1-2):\penalty0 361--367, 2002.

\bibitem[Wang et~al.(2015)Wang, Wang, Zheng, Ducrot, Yodh, Weck, and Pine]{wang2015crystallization}
Yu~Wang, Yufeng Wang, Xiaolong Zheng, {\'E}tienne Ducrot, Jeremy~S Yodh, Marcus Weck, and David~J Pine.
\newblock Crystallization of dna-coated colloids.
\newblock \emph{Nature communications}, 6\penalty0 (1):\penalty0 7253, 2015.

\bibitem[Pusey and Van~Megen(1986)]{pusey1986phase}
Peter~N Pusey and W~Van~Megen.
\newblock Phase behaviour of concentrated suspensions of nearly hard colloidal spheres.
\newblock \emph{Nature}, 320\penalty0 (6060):\penalty0 340--342, 1986.

\bibitem[Anderson and Lekkerkerker(2002)]{anderson2002insights}
Valerie~J Anderson and Henk~NW Lekkerkerker.
\newblock Insights into phase transition kinetics from colloid science.
\newblock \emph{Nature}, 416\penalty0 (6883):\penalty0 811--815, 2002.

\bibitem[McMullen et~al.(2018)McMullen, Holmes-Cerfon, Sciortino, Grosberg, and Brujic]{mcmullen2018freely}
Angus McMullen, Miranda Holmes-Cerfon, Francesco Sciortino, Alexander~Y Grosberg, and Jasna Brujic.
\newblock Freely jointed polymers made of droplets.
\newblock \emph{Physical review letters}, 121\penalty0 (13):\penalty0 138002, 2018.

\bibitem[Luo et~al.(2017)Luo, Smith, Wu, Kim, Ou, and Chen]{luo2017polymerization}
Binbin Luo, John~W Smith, Zixuan Wu, Juyeong Kim, Zihao Ou, and Qian Chen.
\newblock Polymerization-like co-assembly of silver nanoplates and patchy spheres.
\newblock \emph{ACS nano}, 11\penalty0 (8):\penalty0 7626--7633, 2017.

\bibitem[Cai et~al.(2021)Cai, Li, Ravaine, He, Song, Yin, Zheng, Teng, and Zhang]{cai2021colloidal}
Zhongyu Cai, Zhiwei Li, Serge Ravaine, Mingxin He, Yanlin Song, Yadong Yin, Hanbin Zheng, Jinghua Teng, and AO~Zhang.
\newblock From colloidal particles to photonic crystals: Advances in self-assembly and their emerging applications.
\newblock \emph{Chemical Society Reviews}, 50\penalty0 (10):\penalty0 5898--5951, 2021.

\bibitem[Zhang et~al.(2011)Zhang, Wang, Luo, Tikhonov, Kornienko, and Asher]{zhang20112}
Jian-Tao Zhang, Luling Wang, Jia Luo, Alexander Tikhonov, Nikolay Kornienko, and Sanford~A Asher.
\newblock 2-d array photonic crystal sensing motif.
\newblock \emph{Journal of the American Chemical Society}, 133\penalty0 (24):\penalty0 9152--9155, 2011.

\bibitem[Kuo and Lin(2013)]{kuo2013accelerated}
Yung-Chih Kuo and Ching-Chun Lin.
\newblock Accelerated nerve regeneration using induced pluripotent stem cells in chitin--chitosan--gelatin scaffolds with inverted colloidal crystal geometry.
\newblock \emph{Colloids and Surfaces B: Biointerfaces}, 103:\penalty0 595--600, 2013.

\bibitem[Li et~al.(2021)Li, Fan, and Yin]{li2021colloidal}
Zhiwei Li, Qingsong Fan, and Yadong Yin.
\newblock Colloidal self-assembly approaches to smart nanostructured materials.
\newblock \emph{Chemical reviews}, 122\penalty0 (5):\penalty0 4976--5067, 2021.

\bibitem[Ge and Yin(2011)]{ge2011responsive}
Jianping Ge and Yadong Yin.
\newblock Responsive photonic crystals.
\newblock \emph{Angewandte Chemie International Edition}, 50\penalty0 (7):\penalty0 1492--1522, 2011.

\bibitem[He et~al.(2021)He, Gales, Shen, Kim, and Pine]{he2021colloidal}
Mingxin He, Johnathon~P Gales, Xinhang Shen, Min~Jae Kim, and David~J Pine.
\newblock Colloidal particles with triangular patches.
\newblock \emph{Langmuir}, 37\penalty0 (23):\penalty0 7246--7253, 2021.

\bibitem[Sacanna et~al.(2013)Sacanna, Pine, and Yi]{sacanna_engineering_2013}
Stefano Sacanna, David~J. Pine, and Gi-Ra Yi.
\newblock Engineering shape: the novel geometries of colloidal self-assembly.
\newblock \emph{Soft Matter}, 9\penalty0 (34):\penalty0 8096, 2013.
\newblock ISSN 1744-683X, 1744-6848.
\newblock \doi{10.1039/c3sm50500f}.
\newblock URL \url{http://xlink.rsc.org/?DOI=c3sm50500f}.

\bibitem[Sacanna et~al.(2010)Sacanna, Irvine, Chaikin, and Pine]{sacanna2010lock}
Stefano Sacanna, William~TM Irvine, Paul~M Chaikin, and David~J Pine.
\newblock Lock and key colloids.
\newblock \emph{Nature}, 464\penalty0 (7288):\penalty0 575--578, 2010.

\bibitem[Tigges and Walther(2016{\natexlab{a}})]{tigges2016hierarchical}
Thomas Tigges and Andreas Walther.
\newblock Hierarchical self-assembly of 3d-printed lock-and-key colloids through shape recognition.
\newblock \emph{Angewandte Chemie International Edition}, 55\penalty0 (37):\penalty0 11261--11265, 2016{\natexlab{a}}.

\bibitem[Bharti et~al.(2016)Bharti, Kogler, Hall, Klapp, and Velev]{bharti2016multidirectional}
Bhuvnesh Bharti, Florian Kogler, Carol~K Hall, Sabine~HL Klapp, and Orlin~D Velev.
\newblock Multidirectional colloidal assembly in concurrent electric and magnetic fields.
\newblock \emph{Soft Matter}, 12\penalty0 (37):\penalty0 7747--7758, 2016.

\bibitem[Van~Blaaderen(2004)]{van2004colloids}
Alfons Van~Blaaderen.
\newblock Colloids under external control.
\newblock \emph{Mrs Bulletin}, 29\penalty0 (2):\penalty0 85--90, 2004.

\bibitem[Dobnikar et~al.(2013)Dobnikar, Snezhko, and Yethiraj]{dobnikar2013emergent}
Jure Dobnikar, Alexey Snezhko, and Anand Yethiraj.
\newblock Emergent colloidal dynamics in electromagnetic fields.
\newblock \emph{Soft Matter}, 9\penalty0 (14):\penalty0 3693--3704, 2013.

\bibitem[Li et~al.(2020)Li, Palis, M{\'e}rindol, Majimel, Ravaine, and Duguet]{li2020colloidal}
Weiya Li, Herv{\'e} Palis, R{\'e}mi M{\'e}rindol, J{\'e}r{\^o}me Majimel, Serge Ravaine, and Etienne Duguet.
\newblock Colloidal molecules and patchy particles: Complementary concepts, synthesis and self-assembly.
\newblock \emph{Chemical Society Reviews}, 49\penalty0 (6):\penalty0 1955--1976, 2020.

\bibitem[Asakura and Oosawa(1954)]{asakura1954interaction}
Sho Asakura and Fumio Oosawa.
\newblock On interaction between two bodies immersed in a solution of macromolecules.
\newblock \emph{The Journal of chemical physics}, 22\penalty0 (7):\penalty0 1255--1256, 1954.

\bibitem[Col{\'o}n-Mel{\'e}ndez et~al.(2015)Col{\'o}n-Mel{\'e}ndez, Beltran-Villegas, Van~Anders, Liu, Spellings, Sacanna, Pine, Glotzer, Larson, and Solomon]{colon2015binding}
Laura Col{\'o}n-Mel{\'e}ndez, Daniel~J Beltran-Villegas, Greg Van~Anders, Jun Liu, Matthew Spellings, Stefano Sacanna, David~J Pine, Sharon~C Glotzer, Ronald~G Larson, and Michael~J Solomon.
\newblock Binding kinetics of lock and key colloids.
\newblock \emph{The Journal of chemical physics}, 142\penalty0 (17), 2015.

\bibitem[Ruths et~al.(1996)Ruths, Yoshizawa, Fetters, and Israelachvili]{ruths1996depletion}
Marina Ruths, Hisae Yoshizawa, Lewis~J Fetters, and Jacob~N Israelachvili.
\newblock Depletion attraction versus steric repulsion in a system of weakly adsorbing polymer effects of concentration and adsorption conditions.
\newblock \emph{Macromolecules}, 29\penalty0 (22):\penalty0 7193--7203, 1996.

\bibitem[Kuhl et~al.(1998)Kuhl, Berman, Hui, and Israelachvili]{kuhl1998part}
Tonya~L Kuhl, Alan~D Berman, Sek~Wen Hui, and Jacob~N Israelachvili.
\newblock Part 1. direct measurement of depletion attraction and thin film viscosity between lipid bilayers in aqueous polyethylene glycol solutions.
\newblock \emph{Macromolecules}, 31\penalty0 (23):\penalty0 8250--8257, 1998.

\bibitem[Crocker et~al.(1999)Crocker, Matteo, Dinsmore, and Yodh]{crocker1999entropic}
John~C Crocker, Joseph~A Matteo, Anthony~D Dinsmore, and Arjun~G Yodh.
\newblock Entropic attraction and repulsion in binary colloids probed with a line optical tweezer.
\newblock \emph{Physical review letters}, 82\penalty0 (21):\penalty0 4352, 1999.

\bibitem[Rudhardt et~al.(1998)Rudhardt, Bechinger, and Leiderer]{rudhardt1998direct}
Daniel Rudhardt, Clemens Bechinger, and Paul Leiderer.
\newblock Direct measurement of depletion potentials in mixtures of colloids and nonionic polymers.
\newblock \emph{Physical review letters}, 81\penalty0 (6):\penalty0 1330, 1998.

\bibitem[Edwards and Bevan(2012)]{edwards2012depletion}
Tara~D Edwards and Michael~A Bevan.
\newblock Depletion-mediated potentials and phase behavior for micelles, macromolecules, nanoparticles, and hydrogel particles.
\newblock \emph{Langmuir}, 28\penalty0 (39):\penalty0 13816--13823, 2012.

\bibitem[Tigges et~al.(2016)Tigges, Heuser, Tiwari, and Walther]{tigges20163d}
Thomas Tigges, Thomas Heuser, Rahul Tiwari, and Andreas Walther.
\newblock 3d dna origami cuboids as monodisperse patchy nanoparticles for switchable hierarchical self-assembly.
\newblock \emph{Nano letters}, 16\penalty0 (12):\penalty0 7870--7874, 2016.

\bibitem[Brown et~al.(2000)Brown, Smith, and Rennie]{brown2000fabricating}
ABD Brown, CG~Smith, and AR~Rennie.
\newblock Fabricating colloidal particles with photolithography and their interactions at an air-water interface.
\newblock \emph{Physical Review E}, 62\penalty0 (1):\penalty0 951, 2000.

\bibitem[Merkel et~al.(2010)Merkel, Herlihy, Nunes, Orgel, Rolland, and DeSimone]{merkel2010scalable}
Timothy~J Merkel, Kevin~P Herlihy, Janine Nunes, Ryan~M Orgel, Jason~P Rolland, and Joseph~M DeSimone.
\newblock Scalable, shape-specific, top-down fabrication methods for the synthesis of engineered colloidal particles.
\newblock \emph{Langmuir}, 26\penalty0 (16):\penalty0 13086--13096, 2010.

\bibitem[Hernandez and Mason(2007{\natexlab{a}})]{hernandez2007colloidal}
Carlos~J Hernandez and Thomas~G Mason.
\newblock Colloidal alphabet soup: Monodisperse dispersion of shape-designed lithoparticles.
\newblock \emph{The Journal of Physical Chemistry C}, 111\penalty0 (12):\penalty0 4477--4480, 2007{\natexlab{a}}.

\bibitem[Moon et~al.(2007)Moon, Kim, Crocker, and Yang]{moon2007high}
Jun~H Moon, Anthony~J Kim, John~C Crocker, and Shu Yang.
\newblock High-throughput synthesis of anisotropic colloids via holographic lithography.
\newblock \emph{Advanced Materials}, 19\penalty0 (18):\penalty0 2508--2512, 2007.

\bibitem[Jang et~al.(2007)Jang, Ullal, Kooi, Koh, and Thomas]{jang2007shape}
Ji-Hyun Jang, Chaitanya~K Ullal, Steven~E Kooi, Koh, and Edwin~L Thomas.
\newblock Shape control of multivalent 3d colloidal particles via interference lithography.
\newblock \emph{Nano letters}, 7\penalty0 (3):\penalty0 647--651, 2007.

\bibitem[Doherty et~al.(2020)Doherty, Varkevisser, Teunisse, Hoecht, Ketzetzi, Ouhajji, and Kraft]{doherty2020catalytically}
Rachel~P Doherty, Thijs Varkevisser, Margot Teunisse, Jonas Hoecht, Stefania Ketzetzi, Samia Ouhajji, and Daniela~J Kraft.
\newblock Catalytically propelled 3d printed colloidal microswimmers.
\newblock \emph{Soft Matter}, 16\penalty0 (46):\penalty0 10463--10469, 2020.

\bibitem[Tigges and Walther(2016{\natexlab{b}})]{tigges_hierarchical_2016}
Thomas Tigges and Andreas Walther.
\newblock Hierarchical {Self}‐{Assembly} of {3D}‐{Printed} {Lock}‐and‐{Key} {Colloids} through {Shape} {Recognition}.
\newblock \emph{Angewandte Chemie International Edition}, 55\penalty0 (37):\penalty0 11261--11265, September 2016{\natexlab{b}}.
\newblock ISSN 1433-7851, 1521-3773.
\newblock \doi{10.1002/anie.201604553}.
\newblock URL \url{https://onlinelibrary.wiley.com/doi/10.1002/anie.201604553}.

\bibitem[Hernandez and Mason(2007{\natexlab{b}})]{HernandezAlphabet}
Carlos~J. Hernandez and Thomas~G. Mason.
\newblock Colloidal alphabet soup:  monodisperse dispersions of shape-designed lithoparticles.
\newblock \emph{The Journal of Physical Chemistry C}, 111\penalty0 (12):\penalty0 4477--4480, 2007{\natexlab{b}}.
\newblock \doi{10.1021/jp0672095}.
\newblock URL \url{https://doi.org/10.1021/jp0672095}.

\bibitem[Tavacoli et~al.(2013)Tavacoli, Bauër, Fermigier, Bartolo, Heuvingh, and du~Roure]{Tavacoli2013}
Joe~W. Tavacoli, Pierre Bauër, Marc Fermigier, Denis Bartolo, Julien Heuvingh, and Olivia du~Roure.
\newblock The fabrication and directed self-assembly of micron-sized superparamagnetic non-spherical particles.
\newblock \emph{Soft Matter}, 9:\penalty0 9103--9110, 2013.
\newblock \doi{10.1039/C3SM51589C}.
\newblock URL \url{http://dx.doi.org/10.1039/C3SM51589C}.

\bibitem[Saraswat et~al.(2020)Saraswat, Ibis, Rossi, Sasso, Eral, and Fanzio]{Saraswat2020}
Yug~C. Saraswat, Fatma Ibis, Laura Rossi, Luigi Sasso, Huseyin~Burak Eral, and Paola Fanzio.
\newblock Shape anisotropic colloidal particle fabrication using 2-photon polymerization.
\newblock \emph{Journal of Colloid and Interface Science}, 564:\penalty0 43--51, 2020.
\newblock ISSN 0021-9797.
\newblock \doi{https://doi.org/10.1016/j.jcis.2019.12.035}.
\newblock URL \url{https://www.sciencedirect.com/science/article/pii/S0021979719314948}.

\bibitem[Rossi et~al.(2011)Rossi, Sacanna, Irvine, Chaikin, Pine, and Philipse]{rossi_cubic_2011}
Laura Rossi, Stefano Sacanna, William T.~M. Irvine, Paul~M. Chaikin, David~J. Pine, and Albert~P. Philipse.
\newblock Cubic crystals from cubic colloids.
\newblock \emph{Soft Matter}, 7\penalty0 (9):\penalty0 4139--4142, 2011.
\newblock ISSN 1744-683X, 1744-6848.
\newblock \doi{10.1039/C0SM01246G}.
\newblock URL \url{http://xlink.rsc.org/?DOI=C0SM01246G}.

\end{thebibliography}


\section*{Supplementary material}

# Supporting Information
## Specific yet transient bonds between anisotropic colloids

M Mayarani[1,2‡*], Justine Laurent[3], Martin Lenz[1,2], Olivia du Roure[1*] and Julien Heuvingh[1]

[1] PMMH, CNRS, ESPCI Paris, PSL University, Sorbonne Université, Université Paris-Cité, 75005, Paris, France
[2] Université Paris-Saclay, CNRS, LPTMS, 91405, Orsay, France
[3] Gulliver, CNRS, ESPCI Paris, PSL University, Sorbonne Université, 75005, Paris, France.
[‡] Present address: Department of Physics, IIT Palakkad, India, 678623.
Correspondence emails: mayarani@iitpkd.ac.in , olivia.duroure@espci.psl.eu

September 25, 2024

## 3D printing of colloids

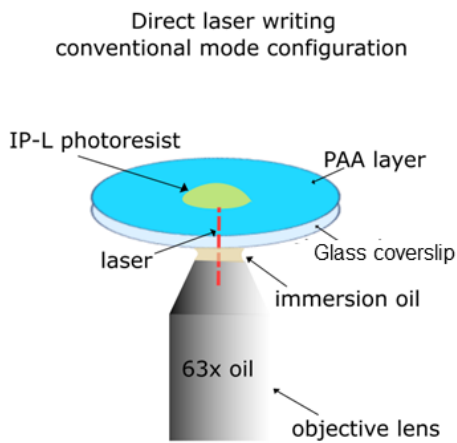

Figure S1: Schematic representation of the arrangement of the objective lens, the substrate and the resin droplet in conventional mode of direct laser writing

Figure S1 represents the arrangement of the microscope objective, glass cover slip substrate, and photoresist in the conventional mode of direct laser writing used in our experiments. It may be noted that the lower working distance of the 63x objective used here necessitates the use of a thin cover slip as substrate for colloidal fabrication. The photoresist is placed above the cover slip and an oil droplet is used between the substrate and the objective of the microscope to increase the numerical aperture of the objective lens. After laser exposure, during the development stage, both the oil and the unreacted resin are washed away by the developer solvent, leaving behind only the printed structures on the substrate.

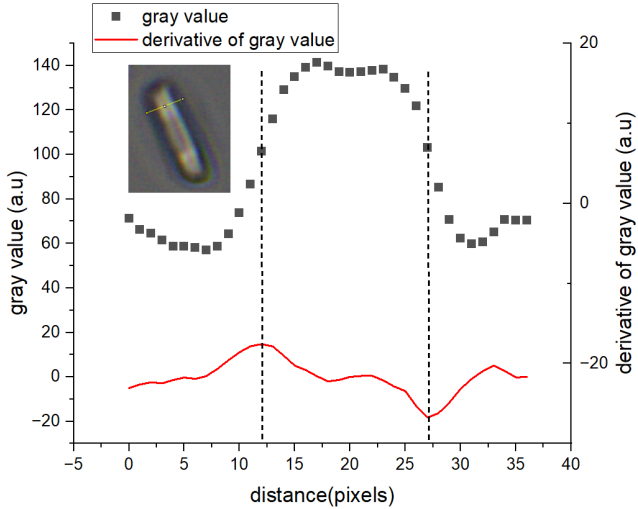

Figure S2: Gray scale profile (black dots) and its derivative along a line drawn across the width of a printed particle obtained from an optical microscopy image, with the particle lying on the substrate with its edge facing down. The image used for analysis is shown in the inset. The inflexion points, denoted by the maxima and minima of the derivative of gray scale profile represents the particle's edges. The distance between the two points gives the width of particle in pixels. In this picture, 20.8 pixels corresponds to $1\,\mu m$.

## Measurement of particle height using optical microscopy

For measuring the height of the printed particles, we employ image analysis using ImageJ software. Image frames are collected from samples printed at different trials. During the assembly of particles, some particles may turn side-wise, enabling the measurement of their height. In the inset of figure S2, an image of a half-disk colloid at a similar instance, with its flat edge parallel to the substrate is shown. Using ImageJ, the gray scale profile along a line drawn across the particle's height is generated (see fig. S2 black dotted curve). The points of inflexion of the line profile are identified from its derivative. The distance between the two inflexion points on the two sides is measured, as the height of half-disk particle. The distances measured in pixels are converted to micrometer based on available calibration for the combination of objective, camera and microscope used. Several measurements were done on particles printed in various attempts and the average of all the measurements is found to be 0.82 $\pm 0.06\,\mu m$. The error is the standard deviation in the average particle height between 5 independent printing trials. The average height of particles at each of the 5 printing trials are calculated from at least 10 different particles from each trial.

## Detection of colloids using ImageJ

Image analysis software ImageJ is used for detecting the position and alignment of the semi-circular colloids. A hole is incorporated at the center of the printed colloids to enable easy detection using ImageJ. A microscopy image of the printed colloid and the elliptical fit performed using ImageJ to the colloidal particle is shown is figure S2. The elliptical fit is carried out to obtain the position of the center of the ellipse(the colloid) and the length of the major axis of the ellipse and the orientation of the major axis with the horizontal axis. These parameters are used to calculate the x-offset values for bonds formed in flat-flat configuration.

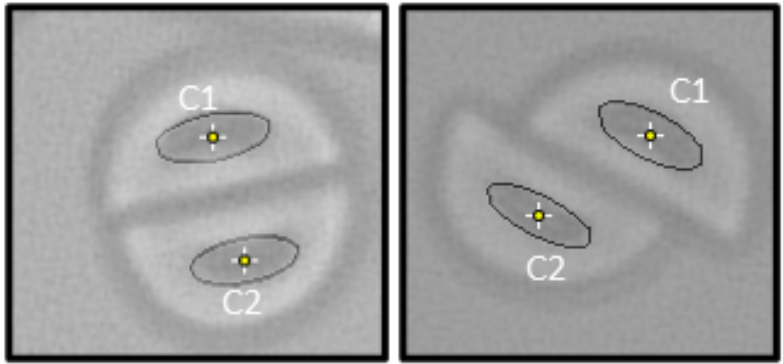

Figure S3: Microscopy image of the semi-circular colloid superimposed with the elliptical fit, and the center of ellipses (C1 and C2) obtained through ImageJ corresponding to two different configurations of the flat-flat bond.

## Experimental chamber design

In order to enable *in-situ* observation of the formation of depletion mediated bonds and their time evolution under 100x magnification, using an oil-immersion objective, a simple experimental chamber is made on the glass substrate containing the printed colloids as shown in figure S3. A rubber O-ring is fixed on the cover slip with the help of a layer of silicon oil underneath the ring. Aqueous depletant solution of required concentration is administrated gently from the top of the chamber, after which it is sealed with a thin glass cover slip of 1 cm diameter. The top glass cover slip assists in avoiding evaporation of the depletant solution during observation.

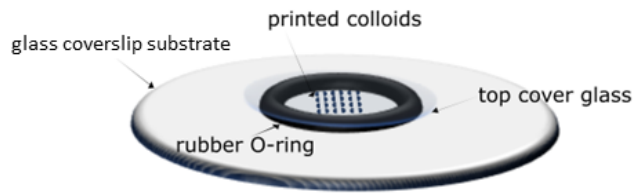

Figure S4: Schematics showing the experimental chamber used for observing the self-assembly of printed colloids. The schematic is not to scale: the glass cover slip is of 30mm diameter, the O-ring is roughly 5mm in diameter while the printed zone is typically of the order of $200\mu$m x $200\mu$m.

## Diffusion of colloidal particles

As the printed particles are released from their initial position by dissolving the sacrificial layer, they start diffusing along the glass surface. The pre-positioning of the particles is lost soon after the introduction of the depletant solution. The angle of orientation of all the particle in the image frame with the positive x axis is plotted at different times. Before placing the depletant, at $t = 0$, all the particles are aligned in the same direction. Soon after the introduction of the depletant solution followed by the dissolution of PAA layer, the initial particle arrangement is completely lost leading to random orientation of particles as evident from figure S5.

Using an optical microscope equipped with a 100x objective and a MIchrome 6 camera with CMOS sensor, movies of the particles are captured at a frame rate of 1 per second. Using ImageJ, the movie frames are analysed to deduce the x and y co-ordinates of the centroid of the particles. The trajectories of the particles are created from the time series images of the particles fluctuating under thermal agitation and analysed to obtain the mean square displacement(MSD) as a function of

3

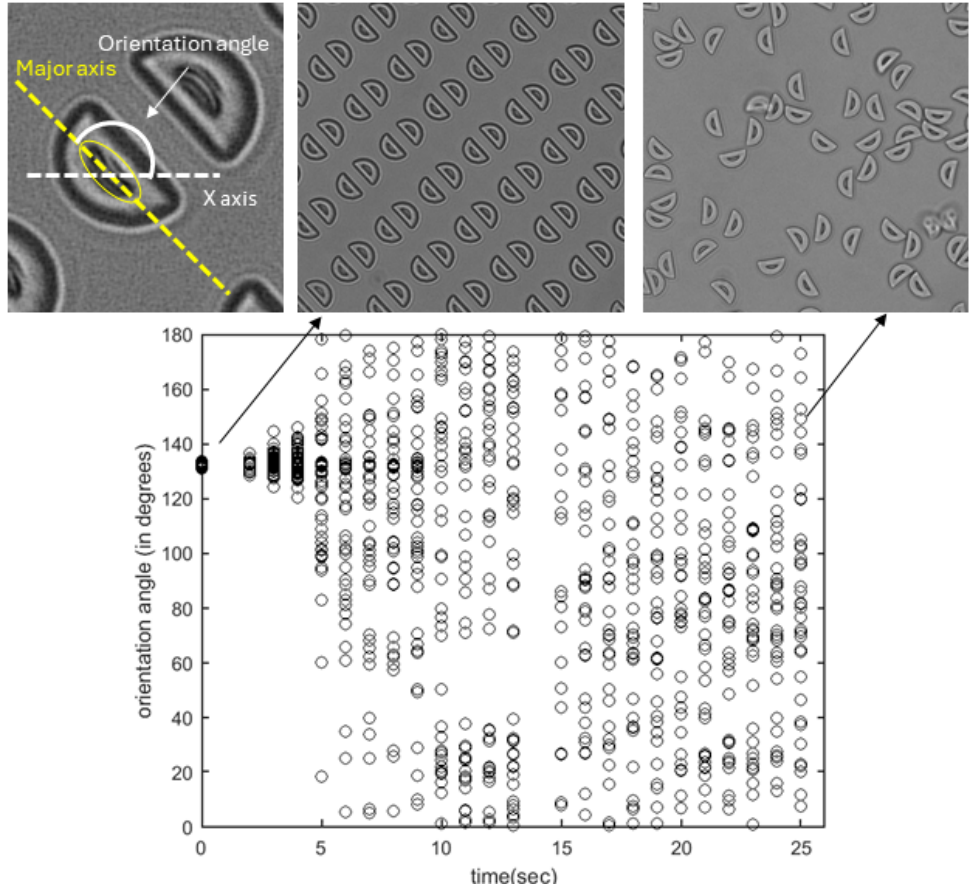

Figure S5: Angle of orientation of the particles with the positive x axis, at different instances during the self-assembly process.

delay time $t$ according to eq.1. From the slope of MSD Vs. delay time plot, the diffusion co-efficient, D of the particles is deduced, assuming that the diffusion is taking place in a 2 dimensional space (see Fig.S6).

$$MSD = <[r(t+\tau) - r(t)]^2> = 4Dt \quad (1)$$

## Analysis of bond breaking

In-order to identify the time at which the bonds between two colloids in flat-flat configuration break apart, we measure the angle $\theta$ between the major axes of the two elliptical fits to the colloidal particles. When the colloids are bound, $\theta$ remains close to $0°$. A sudden jump in the angle between the ellipses indicate a breaking event. In figure 4, the colloids are bound between time 0-36sec and abruptly breaks apart at 36sec, which is indicated by a sudden increase in $\theta$.

## Calculation of change in excluded volume

The half-disk colloids harbour an exclusion layer of thickness $\delta$ around their surface. When two half-disks are in contact through their flat faces, the gain in excluded volume is simply $V = 2R\delta h$ for each half disk of radius $R$ and height $h$, for a total of

$$\Delta V_{f-f} = 4R\delta h.$$

4

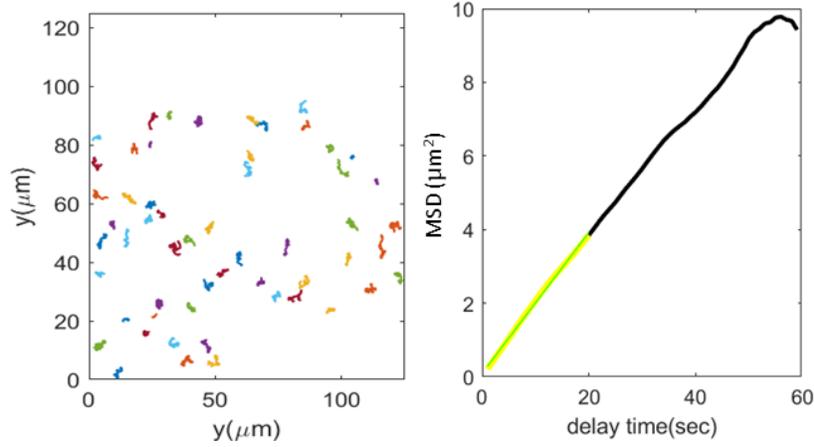

Figure S6: (a) Trajectories of printed particles along the glass surface soon after the introduction of depletant solution inside the experimental chamber. (b) Mean Square Displacement Vs. delay time plot of the particles for a total duration of 1 minute. The data is fitted with a straight line (shown in green) at the initial times (highlighted in yellow).

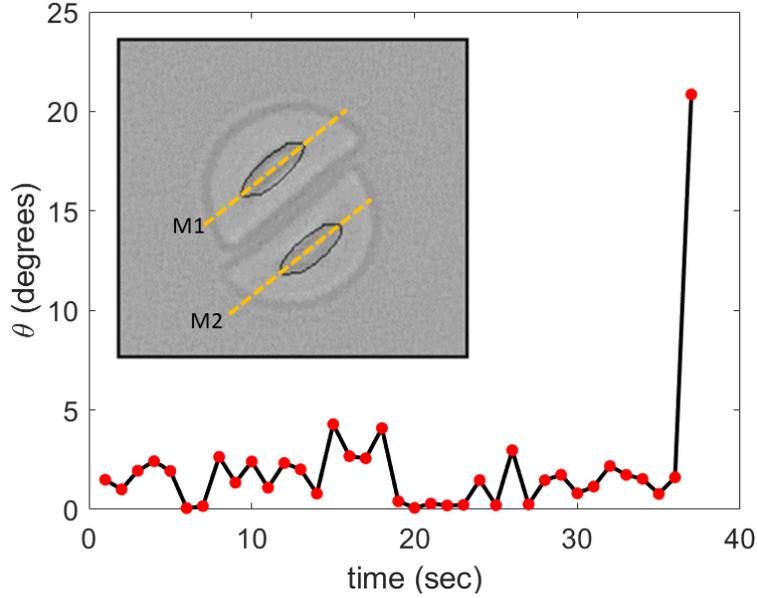

Figure S7: Time evolution of the angle $\theta$ between the major axes of two elliptical fits to the half circle colloids in the bound stage. $\theta$ remains close to zero while the colloids are in bound state. Bond breaking is characterized by abrupt jump in $\theta$. The inset shows the elliptical fits to two bound colloids and the major axes of the two ellipses.

For two round sides in contact, one must calculate the area of the circular segment of radius $R+\delta$ and sagitta $\delta$. This can be done by calculating the circular sector and subtracting the triangular portion

$$S_1 = (R+\delta)^2 \theta_1 - R((R+\delta)^2 - R^2)^{1/2}$$

with $\theta_1 = acos(\frac{R}{R+\delta})$. The excluded volume gain for each half-disks is $hS_1$, for a total of

$$\Delta V_{r-r} = 2hS_1.$$

5

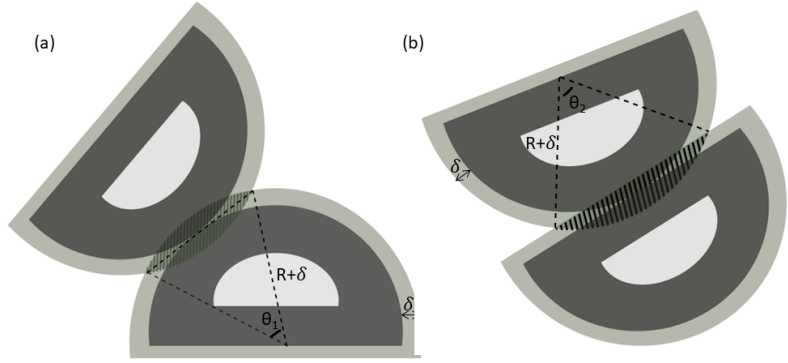

Figure S8: Schematics showing the excluded volume around the half-disk colloids and the change in excluded volume on Round-Round close contact.

For a round side in contact with a flat face, the projection of the excluded volume difference will be a circular segment of radius $R + \delta$ and sagitta $2\delta$. Its surface is

$$S_2 = (R+\delta)^2 \theta_2 - (R-\delta)((R+\delta)^2 - (R-\delta)^2)^{1/2}$$

with $\theta_2 = acos(\frac{R-\delta}{R+\delta})$. The excluded volume gain is in this case

$$\Delta V_{f-r} = hS_2.$$

The depletant exclusion thickness was taken as $\delta = \frac{2R_g}{\sqrt{\pi}}$[1], with $R_g$ the radius of gyration of the polymer. We took $R_g = 50.2 \pm 1.5 nm$ for PEG 600kDa[2].

From the relative errors of $R$ (1%), $h$ (7%) and $\delta$ (3%), we computed an error 8.3% for $V_{f-r}$ and $V_{r-r}$ and 7.6% for $V_{f-f}$.

6

# Fluctuations and bonds breaking at different concentrations

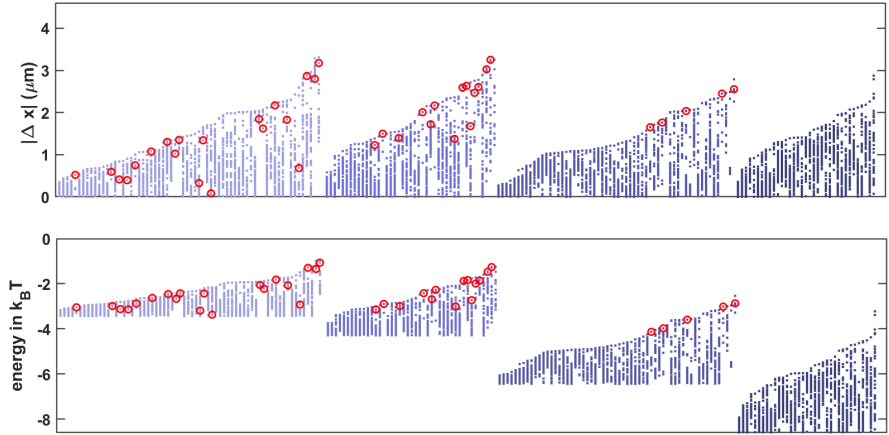

Figure S9: Spectrum of $|\Delta x|$ values (top panel) and the corresponding bond energies (bottom panel) explored by the colloidal pairs in flat-flat configuration. The $|\Delta x|$ and bond energy values explored by each pair of colloids during its temporal evolution is plotted vertically on the respective panels. The pairs are indexed in the ascending order of highest value of $|\Delta x|$ among the various values they explore. Four different shades of blue represent $|\Delta x|$ values and bond energies extracted from the systems with four different concentrations of depletants, viz 0.008 mg/ml, 0.01 mg/ml, 0.015mg/ml and 0.02mg/ml in order from left to right. The corresponding minimum of the potential wells are at $-3.43 k_B T$, $-4.29 k_B T$, $-6.44 k_B T$, and $-8.59 k_B T$ respectively. The red open circles corresponds to the point at which bond breaking takes place.

# References

[1] Gerard J. Fleer, Alexander M. Skvortsov, and Remco Tuinier. Mean-Field Equation for the Depletion Thickness. *Macromolecules*, 36(20):7857–7872, October 2003.

[2] K Devanand and JC Selser. Asymptotic behavior and long-range interactions in aqueous solutions of poly (ethylene oxide). *Macromolecules*, 24(22):5943–5947, 1991.


\end{document}